\documentclass[conference]{IEEEtran}

\usepackage{epsfig,endnotes}

\usepackage{makecell}
\usepackage{tcolorbox}
\usepackage{enumitem}
\usepackage{blindtext}
\usepackage{tabularx}
\usepackage{booktabs}
\usepackage[dvipsnames]{xcolor}
\usepackage{hyperref}
\usepackage{amssymb}
\usepackage{multirow}

\definecolor{customblue}{HTML}{006ca6}
\definecolor{customgreen}{HTML}{009264}
\definecolor{custombrown}{HTML}{ff3d00}
\AtEndPreamble{

 \hypersetup{
  colorlinks = true,
  linkcolor = customblue,
  anchorcolor = purple,
  citecolor = customgreen,
  filecolor = purple,
  urlcolor = custombrown
 }
}

\usepackage{tikz}
\usetikzlibrary{positioning,arrows.meta}

\usepackage{cite}

\usepackage[skip=\baselineskip]{caption}

\newtcbox{\roundedtrend}{
  on line,
  boxsep=1pt,
  left=1pt,
  right=1pt,
  top=0.25pt,
  bottom=0.25pt,
  arc=2pt,
  boxrule=0pt,
  colback=blue!20,
  colframe=blue!20
}

\newcommand{\trend}[1]{%
  \roundedtrend{\textbf{T#1}}%
}

\newtcbox{\roundedopp}{
  on line,
  boxsep=1pt,
  left=1pt,
  right=1pt,
  top=0.25pt,
  bottom=0.25pt,
  arc=2pt,
  boxrule=0pt,
  colback=red!25,
  colframe=red!25
}

\newcommand{\opp}[1]{%
  \roundedopp{\textbf{O#1}}%
}

\newtcbox{\roundeddir}{
  on line,
  boxsep=1pt,
  left=1pt,
  right=1pt,
  top=0.25pt,
  bottom=0.25pt,
  arc=2pt,
  boxrule=0pt,
  colback=green!15,
  colframe=green!15
}

\newcommand{\dir}[1]{%
  \roundeddir{\textbf{D#1}}%
}

\begin{document}

\title{SoK: From Generation to Consumption of \\ Privacy Documents in Software Systems}

\author{\IEEEauthorblockN{Shidong Pan$^*$\thanks{$^*$Shidong Pan completed most of the work when he was a postdoctoral research fellow at Columbia and NYU. }}
	\IEEEauthorblockA{Columbia University, USA\\ 
    New York University, USA\\
	shidong.pan@nyu.edu}
	\and
	\IEEEauthorblockN{Clark LaChance}
	\IEEEauthorblockA{University of Maine, USA\\
		clark.lachance@maine.edu}
	\and
	\IEEEauthorblockN{Zhen Tao}
	\IEEEauthorblockA{ Technical University of Munich,\\
    Germany\\
		zhen.tao@tum.de}
        	\and
	\IEEEauthorblockN{Sepideh Ghanavati}
	\IEEEauthorblockA{University of Maine, USA\\
		sepideh.ghanavati@maine.edu}
}

\IEEEoverridecommandlockouts
\makeatletter\def\@IEEEpubidpullup{6.5\baselineskip}\makeatother
\IEEEpubid{\parbox{\columnwidth}{
		Network and Distributed System Security (NDSS) Symposium 2027\\
		TBD 2027, TBD, USA\\
		ISBN 979-8-9919276-8-0\\  
		https://dx.doi.org/10.14722/ndss.2027.[23$|$24]xxxx\\
		www.ndss-symposium.org
}
\hspace{\columnsep}\makebox[\columnwidth]{}}

\maketitle

\begin{abstract}

Privacy documents (e.g., privacy policies) are a central mechanism through which digital services disclose data practices and seek user consent. 
Over the past decades, research on privacy documents has expanded significantly, encompassing not only traditional privacy policies but also short notices (e.g., privacy labels) and interface-level transparency mechanisms. 
As this research area continues to grow, it has become increasingly difficult to obtain a coherent view of how privacy documents are created, analyzed, evaluated, and maintained across their lifecycle. 
This SoK provides a unified, lifecycle-oriented view of privacy documents from a software engineering perspective. 
We systematically review and analyze 290 papers published between 2010 and 2025, organizing them around five research questions that 
examine how privacy documents are (1) defined and scoped, (2) generated, (3) analyzed and extracted, (4) checked for inconsistencies and noncompliance, and (5) evaluated and improved for usability.
Building on our findings, we identify 15 key research trends and 21 open opportunities. 
We further chart four broader research directions that highlight (i) emerging challenges in AI-centric platforms, (ii) the need for diverse and up-to-date data foundations, (iii) LLM-based unified policy-code analysis, and (iv) dual usability for end-users and developers.
We hope this SoK provides a shared foundation for future research on privacy policies and privacy documents.

\end{abstract}

\section{Introduction}
\label{sec:introduction}

Privacy regulations commonly mandate that software systems provide user-facing disclosures that explain how personal data are collected, used, shared, and protected~\cite{GDPR, CCPA, LGPD, PIPL, APPs}. These disclosures take the form of privacy documents, such as privacy policies, privacy labels, permission rationales, and cookie notices. 
Privacy documents are intended to serve as a central mechanism for transparency, accountability, and user empowerment across modern digital ecosystems~\cite{sadeh2013usable, amos2021privacy, sathyendra2017identifying, mhaidli2023researchers}. 
They also play a critical compliance role, acting as the interface between regulatory requirements, software behavior, and user understanding~\cite{xiang2023policychecker, zhan2024vpvet, zhou2023policycomp, nguyen2021share}.
Despite their importance, privacy documents often fail to achieve their intended goals. Prior work has repeatedly shown that privacy policies are lengthy, vague, and difficult to understand~\cite{obar2020biggest, wagner2023privacy, ermakova2015readability, audich2021improving, becher2020law}. 
More recent formats, such as privacy labels and contextual privacy policies, aim to address these shortcomings, yet they introduce new challenges related to accuracy, consistency, and maintainability~\cite{emami2021informative, windl2022automating, pan2024hope, gong2025towards}. 
Empirical studies further reveal widespread mismatches between privacy documents and actual software behaviors, as well as inconsistencies across different document types, platforms, and jurisdictions~\cite{tokas2022static, surma2024examining, kounoudes2024right, van2022setting}. 

Research on privacy documents has grown substantially over the past two decades and spans multiple disciplines, including computer science (CS), law, public policy, economics, and psychology. 
Within CS, research aims to address a broad range of problems, such as automated generation and natural language analysis of privacy policies, regulatory compliance checking, and usability-driven redesigns of privacy notices. 
However, this body of work remains fragmented, particularly across research areas such as privacy, software engineering, and natural language processing (NLP).
This fragmentation is also reflected in the current survey literature. Existing surveys typically focus on narrow slices of the literature, such as NLP techniques for privacy policies~\cite{Javed2024, adhikari2025natural, mhaidli2023researchers} or usability studies of specific notice designs~\cite{alamri2023privacy, Schaub2015}. 
There is currently no systematization that unifies these efforts across document types, technical approaches, and application goals.

\begin{table*}[ht]
    \centering
    \footnotesize
    \caption{Comparison of related work and our SoK.}
    \renewcommand{\arraystretch}{1.2}
    \resizebox{0.9\textwidth}{!}{
        \begin{tabular}{|l|c|c|c|c|c|c|c|c|}
            \hline
            \textbf{Work} & \textbf{Type} &  \textbf{Venue \& Year} & \textbf{Scope} & \textbf{Generation} & \textbf{Analysis} & \textbf{Compliance} & \textbf{Usability} & \textbf{Papers} \\
            \hline
            Schaub et al.~\cite{Schaub2015} & SoK & SOUPS 2015 & All privacy documents & $\times$ & $\times$ & $\times$ & $\checkmark$ & Unspecified \\
            \hline
            Mhaidli et al.~\cite{mhaidli2023researchers} & Other& PoPETS 2023 &  Privacy policies &  $\times$ & $\checkmark$ & $\times$ & $\times$ & N/A \\
            \hline
            Alamri et al.~\cite{alamri2023privacy} & SLR & CADE 2023 & Privacy policies & $\times$ & $\times$ & $\times$ & $\checkmark$ & Unspecified \\
            \hline
            Javed and Sajid~\cite{Javed2024} & SLR & CSUR 2024 & Privacy policies & $\times$ & $\checkmark$ & Partial & $\times$ & 202 \\
            \hline
            Adhikari et al.~\cite{adhikari2025natural} & SLR & Preprint 2025 & Privacy policies & $\times$ & $\checkmark$ & $\times$ & $\times$ & 109 \\
            \hline
            This Work & SoK & -- & All privacy documents & $\checkmark$ & $\checkmark$ & $\checkmark$ & $\checkmark$ & 290 \\
            \hline
        \end{tabular}
        }
    \label{tab:related_word_comparison}
\end{table*}

This paper presents the first SoK on the full \emph{lifecycle} of privacy documents.
We conceptualize privacy documents as software components and analyze them from a software engineering perspective. 
This lifecycle perspective is important because privacy documents are not static legal artifacts: they are defined, generated, analyzed, validated, and consumed as part of evolving software systems, and these stages are inter-connected.
At the same time, privacy documents are produced and maintained at scale by thousands of developers and service providers, making automation, maintainability, traceability, and quality assurance central engineering challenges.
Therefore, our analysis covers the entire lifecycle of privacy documents, including their generation, analysis, consistency \& compliance, and consumption. 
To ground this SoK, we systematically review 290 papers published between 2010 and 2025 across top-tier venues in security and privacy, software engineering, and NLP.
Specifically, the SoK is organized by the following RQs:

\begin{itemize} [leftmargin = *]
    \item \textbf{RQ1:}\textit{What definitions and scopes of privacy documents are adopted in the literature?}
    
    \item \textbf{RQ2:} \textit{What methods, techniques, and tools are used for privacy document generation?}

    \item \textbf{RQ3:} \textit{What methods, techniques, and tools are used to analyze privacy documents and extract privacy-relevant features?}

    \item \textbf{RQ4:} \textit{What forms of inconsistencies and noncompliance involving privacy documents have been studied?} 

    \item \textbf{RQ5:} \textit{What methods, techniques, and tools are used for evaluating and improving the usability and readability of privacy documents?}
\end{itemize}

Across the five RQs, our SoK identifies 15 research trends and 21 open opportunities across the full lifecycle of privacy documents.
Briefly, privacy policies still dominate the literature, but document formats are evolving, while existing studies remain largely English-centric and focus on mainstream-market (RQ1). 
Privacy document generation remains underexplored and mostly artifact-driven, with limited support for traceability, developer usability, and collaborative workflows (RQ2). 
Analysis is increasingly shaped by NLP and LLM-based methods, although rule-based and manual approaches remain important for structured representation and expert validation (RQ3).
Consistency and compliance studies show widespread inconsistencies, especially across software behavior, privacy policies, and regulations, but remain concentrated on mobile apps and GDPR-focused analyses (RQ4). 
Usability research confirms that privacy documents remain long, vague, and hard to read, while disclosures are gradually becoming more contextual, just-in-time, and actionable (RQ5).

Furthermore, building on the identified insights and gaps, we chart four boarder research directions to guide future privacy document research: 
1) addressing privacy challenges in AI-centric platforms; 
2) developing large-scale, up-to-date, multilingual, multi-source, and longitudinal data foundations; 
3) advancing LLM-based unified analysis of privacy policies and software artifacts; and 
4) improving dual usability for both end-users and developers. 
Collectively, these directions emphasize the need to study privacy documents as evolving software components that must remain accurate, traceable, compliant, and usable throughout their lifecycle.


\section{Related Work}
\label{sec_background}

Prior efforts have studied various facets of privacy document research, specifically, privacy policies~\cite{harkous2018polisis, andow2019policylint, andow2020actions, cui2023poligraph, zhou2023policycomp, amos2021privacy, shipp2020private, wagner2023privacy, adhikari2023evolution, singh2011evaluating}. 
Javed and Sajid's recent systematic review of 202 studies (through 2023) provided a broad survey of privacy policy research, including policy readability and completeness, automated NLP analyses, and end-user comprehension of policies via summarization and visualization tools~\cite{Javed2024}. 
Complementing this, Adhikari et al.~\cite{adhikari2025natural} offered a focused survey on the use of NLP for privacy policies, categorizing work into tasks such as information extraction, classification, summarization, and compliance alignment. 
Meanwhile, researchers' perspectives were explored by Mhaidli et al.~\cite{mhaidli2023researchers}, who reported on the methodological gaps faced when analyzing privacy policies, such as legal ambiguity and tool limitations. 
Similarly, Alamri et al.~\cite{alamri2023privacy} identified key research barriers, including limited cross-disciplinarity and a need for real-world validation.


Table~\ref{tab:related_word_comparison} shows the difference between this and prior work. 
While prior works offer valuable insights into analysis methods and research experiences, they tend to focus on either a single document type (e.g., the privacy policy) or a specific methodological domain (e.g., NLP). 
In contrast, this SoK provides a broader systematization of privacy documents, including policies, labels, icons, and contextual notices, spanning across their entire lifecycle, from generation, analysis to compliance, and usability dimensions, highlighting their interdependencies and identifying open challenges. 
Beyond categorizing individual stages, our cross-stage analysis identifies dependencies that are difficult to observe in stage-specific surveys, as specified in Section~\ref{sec:research_framework}.

\section{Methodology}
\label{sec:methodology}

\begin{figure*}[t]
    \centering
    \includegraphics[width=\linewidth]{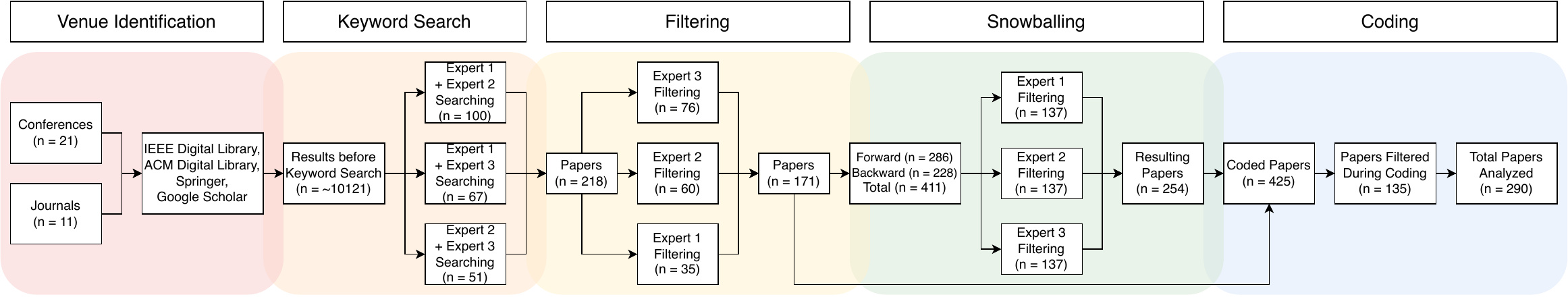}
    \vspace{-15pt}
    \caption{An overview of the paper collection methodology.}
    \label{fig:methodology_flowchart}
\vspace{-10pt}
\end{figure*}

In this section, we describe our methodology for collecting and coding papers.
Fig.~\ref{fig:methodology_flowchart} shows an overview of our methodology and the total paper counts at each stage.
Overall, we followed formal SLR guidance~\cite{kitchenham2009systematic} by using predefined RQs, conducting systematic query-based searching and screening, and inductive open-coding.

\begin{figure}[t]
    \centering
    \includegraphics[width=\linewidth]{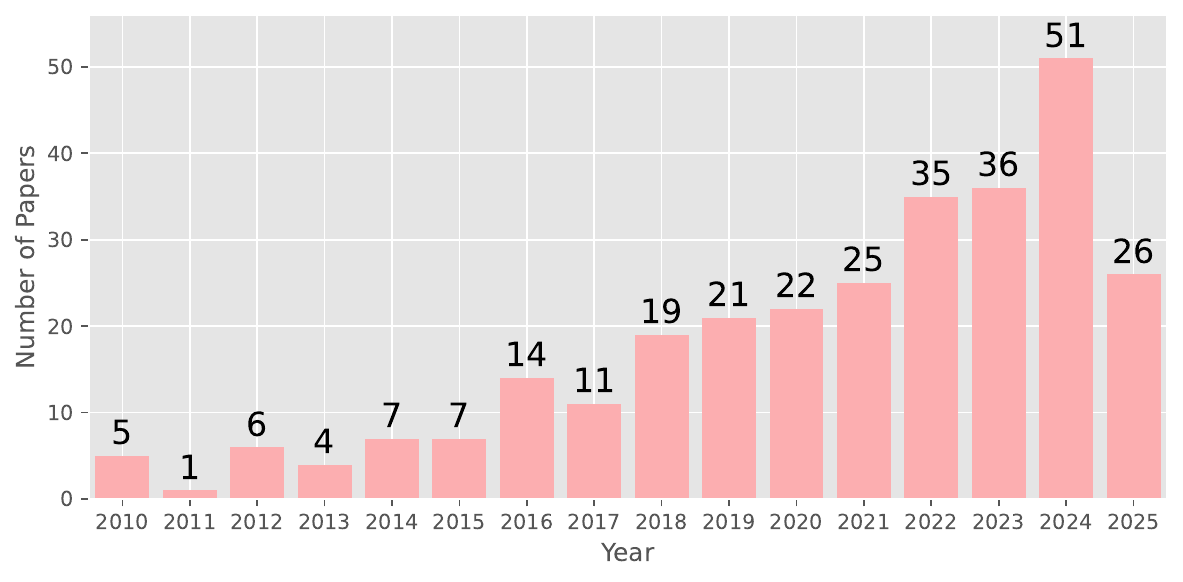}
    \vspace{-15pt}
    \caption{Papers published per year from 2010 to 2025.}
    \label{fig:papersperyear}
\vspace{-10pt}
\end{figure}

\subsection{Data Collection}
\label{sec:data-collection}

\paragraph{Venue Identification}

In the first step, we identified a list of venues that may contain papers about privacy documents based on our RQs. 
As we adopt a broad definition for ``privacy documents'', we examined more than just privacy venues to ensure that our list captures all relevant works. 
We compiled an initial list of 16 privacy/security conferences, 11 software engineering conferences, 5 NLP conferences, and 11 journals in the same relevant fields. 
Each conference is ranked B or higher, as per the CORE ranking~\cite{CORERanking}. 
During venue screening, 11 candidates were excluded because 1) they did not meet our selected CORE-ranking threshold (e.g., Australasia Conference on Information Security and Privacy) or; 2) their scope was highly specialized in a domain with limited expected overlap with the privacy-document lifecycle (e.g., International Symposium on Software Engineering for Adaptive and Self-Managing Systems)
In total, we obtained 32 venues to search. 




\paragraph{Keyword Search}

The first three authors manually collected an initial set of papers from each of the 32 venues based on a keyword search. 
In the first round, the workload was equally distributed, with two authors assigned to search each. 
By ensuring that two authors review each selected conference and journal, we aimed to minimize the total number of missed papers. 
We identified the following keywords to search in the titles and abstracts of all papers in each venue to filter the total list of papers searched: 

\vspace{-3pt}
\begin{tcolorbox}[colback=gray!10,colframe=black!30,boxrule=0.1pt, boxsep=0.1pt]
\vspace{-3pt}

\textit{(privacy AND (polic(y/ies) OR document OR label OR notice OR consent))} \textit{OR} 

\textit{(polic(y/ies) AND (document OR label OR notice OR consent))}
\vspace{-3pt}
\end{tcolorbox}
\vspace{-3pt}

Search terms were derived deductively from our operational definition of privacy documents. ``policy'' and ``policies'' capture conventional long-form privacy policies, while ``document'', ``label'', ``notice'', and ``consent'' capture the broader disclosure formats covered by RQ1. 
Because all retrieved papers subsequently underwent expert screening and the seed corpus was expanded through snowballing, we intentionally adopted a broad query to prioritize recall.

Some venues, such as USENIX Security, did not allow for advanced keyword searches, which resulted in approximately 10,121 raw hits to search, with many false positives. To screen these papers, we instead combined a manual analysis of the retrieved papers together with Google Scholar and venue-restricted DBLP~\cite{dblp} title searches.
Each author also excluded any paper that did not appear to meet the following criteria. Ambiguous cases were included at this step to avoid missing relevant works.

\begin{itemize} [noitemsep,topsep=0pt] 
    \item Papers must be written in English.
    \item Papers must have been published between 2010 - 2025.
    \item Papers are not from a table of contents, an entire proceedings, a tutorial, a book, a standard definition, a workshop paper, a poster, a technical report, or an extended abstract.
    \item The paper must focus on privacy document analysis, methods, tools, models, techniques, and/or datasets. We define a broad definition of privacy documents as any artifact used to communicate data practices and privacy obligations, which includes long-form textual policies, short notices, labels, and other disclosures.
    \item Papers focusing on ``Terms of Service'' were excluded.
\end{itemize}

Although ToS documents often accompany privacy policies, their primary function is to outline contractual relationships and user obligations rather than to describe data privacy practices~\cite{bernhard2025multilingual}. In contrast, privacy documents are created by service providers to communicate their responsibilities and obligations regarding data privacy. 
This conceptual distinction ensures that our analysis focuses specifically on artefacts that aim to inform users about privacy and data governance.

\paragraph{Filtering}
\label{sec:filtering_1}

We extracted 218 papers after completing the initial data collection. To minimize the number of false positives in our set, we conducted an additional round of review at this point. 
Since each venue was checked by two authors, the third author, who did not initially search for a given venue, reviewed any disagreements between the papers collected by the other two authors. 
the third author examined the title and abstract and wrote a short rationale for inclusion or exclusion. All authors then discussed and approved these rationales before finalizing the decisions.
The final decisions were based on whether each paper satisfied the criteria listed above. After this filtering stage, 171 papers remained.



\begin{table*}[t]
\centering
\small
\caption{Summary of research trends, open opportunities, and research directions. 
The numbers (\%) in the \textit{Aspect} column indicate the distribution of papers across lifecycle stages. Counts overlap because papers may address multiple lifecycle stages.}
\label{tab:rq_summary}
\vspace{-10pt}
\resizebox{0.91\textwidth}{!}{
\renewcommand{\arraystretch}{1.25}
\begin{tabularx}{\textwidth}{p{0.08\textwidth} X X}
\toprule
\textbf{Aspect} & \textbf{Research Trends} & \textbf{Open Opportunities} \\
\midrule

\raisebox{-0.7ex}{\textbf{Scope}}
&
\trend{1} Privacy policies dominate; 
\trend{2} Formats are evolving;
\trend{3} Studies remain English- and mainstream-market centric &
\opp{1} Study emerging privacy document formats; \opp{2} Other linguistic and cultural contexts 
\\
\midrule

\raisebox{-2ex}{\textbf{Generation}}
\raisebox{-2ex}{{32 (11.0\%)}}
&
\trend{4} Generation is underexplored; 
\trend{5} Existing generation approaches are mostly artifact-driven; 
\trend{6} Individual human factors are commonly considered during generation process &
\opp{3} Use code evidence for authentic statements; \opp{4} Standardize formats for traceability; \opp{5} Support developer usability; 
\opp{6} Collaborative and dynamic workflows; 
\opp{7} Tension between productivity and responsibility
\\
\midrule

\raisebox{-2ex}{\textbf{Analysis}}
\raisebox{-2ex}{{205 (70.7\%)}}
&
\trend{7} Rule-based methods provide representations; 
\trend{8} Learning-based methods inherit representations; 
\trend{9} NLP advances analysis, but facing explainability and hallucination issues; 
\trend{10} Manual analysis is still mainstream &
\opp{8} Extract subtler privacy practices; 
\opp{9} Expand HiTL approaches; 
\opp{10} Ananlysis based on large-scale expert annotation; 
\opp{11} Longitudinal dataset for evolution; 
\opp{12} Multilingual and multi-source datasets 
\\
\midrule

\raisebox{-2ex}{\textbf{Compliance}}
\raisebox{-2ex}{{130 (44.8\%)}}
&
\trend{11} Studies focus on mobile apps as mature program analysis; 
\trend{12} Studies focus on GDPR; 
\trend{13} Inconsistencies are widespread, especially across policies and labels 
&
\opp{13} Cover AI-centric ecosystems; 
\opp{14} Broader jurisdictions; 
\opp{15} Ambiguity persists in regulations;
\opp{16} Industry standards and platform requirements;
\opp{17} Real-time compliance checking; 
\opp{18} Multilateral inconsistency
\\
\midrule

\raisebox{-1ex}{\textbf{Usability}}
\raisebox{-1ex}{{126 (43.4\%)}}
&
\trend{14} Privacy documents remain long, vague, and hard to read; 
\trend{15} Disclosures are becoming more contextual, just-in-time, and readable 
&
\opp{19} Revisit machine-readable formats in the LLM era; 
\opp{20} From system-centric to user-centric designs;
\opp{21} Actionable privacy controls \\
\midrule
\midrule
\raisebox{0.4ex}{\textbf{\makecell[c]{Research \\ Direction}}}
&
\multicolumn{2}{l}{
\makecell[l]{
\dir{1} AI-centric platforms and emerging privacy challenges, e.g., prompt-centric and high autonomy (\opp{13} \opp{17} \opp{19}) \\
\dir{2} Large-scale, update-to-date, multilingual, multi-source, and longitudinal data foundation (\opp{1} \opp{10} \opp{11} \opp{12} \opp{18})
\\
\dir{3} LLM-based unified policy-software analysis for contextualized and faithful interpretation (\opp{3} \opp{4} \opp{8} \opp{9} \opp{15})
\\
\dir{4} Dual usability for end-users and developers, advocating more usable privacy engineering tools
(\opp{5} \opp{6} \opp{20} \opp{21})
}%
}%

\\

\bottomrule
\end{tabularx}
}%
\vspace{-10pt}
\end{table*}

\paragraph{Forward and Backward Snowballing}

For each paper, the authors performed a single round of forward snowballing \cite{snowballing_guidelines} using Google Scholar and one round of backward snowballing based on the cited references. 
They applied most of the same inclusion criteria, including the same keyword filtering approach, which yielded 286 additional papers for forward snowballing and 228 additional papers for backward snowballing.
To improve recall beyond the bounded seed search, we relaxed two source-level restrictions during snowballing: candidate papers were not limited to the 32 seed venues, and workshop papers were eligible.
Since each author performed the snowballing independently and the year limit varied between forward and backward snowballing, some papers were collected more than once. 
After removing these duplicates, 411 unique papers were obtained via snowballing.


To verify that all relevant papers were identified and that we reached saturation after the snowballing rounds, we randomly selected a sample of 10\% of the works obtained from snowballing to perform another round. This process did not result in additional papers that fit our inclusion criteria and were not already in our dataset, indicating that saturation was reached. Therefore, we do not believe that a full second round of snowballing is required.

We then applied a similar filtering technique to these 411 papers to ensure that all included papers were verified by at least 2 authors. Each author reviewed another author's list of papers to provide a second opinion on relevance. This resulted in the removal of 157 papers, leaving 254 unique papers from the snowballing process. 
Combining with our initial set of 171 papers, we had 425 papers to manually code.

\subsection{Coding and SoK-ization}
\label{sec:coding}

We followed an open-coding procedure to construct our codebook~\cite{huberman2014qualitative, mayring2021qualitative}. 
The first three authors, as annotators, independently coded the same initial 20 papers.
For each of our research questions, annotators were tasked to provide an initial list of high-level categories (L1) and individual codes (L2) for more details, with a small set of initial codes added prior to coding. 
The annotators compared their independently generated categories and codes, resolved differences through discussion, and consolidated the final codebook.
Then, the remaining papers were roughly equally distributed to annotators and they coded them independently.
After completing the rest of the coding, all authors met and discussed the edge cases to ensure consistency and coherence of the codes. 
In this stage, the annotators also reviewed the content of the papers; thus, some papers were identified as irrelevant. 
For example, some papers initially appeared to directly analyze ``privacy policies'', but are actually in the context of security policies, which are out of our scope. 
In total, 135 papers were ultimately determined to be unrelated to privacy documents, resulting in a final count of 290 papers analyzed within this SoK. 
Fig.~\ref{fig:papersperyear} shows the numbers of papers published per year between 2010 and 2025, showing a steady increase over time.



Throughout this section, we enumerate research trends as \trend\#, opportunities for future work as \opp\#, and broader research directions as \dir\#, to facilitate visual presentation.
Table~\ref{tab:rq_summary} presents a summary of this SoK.
The uneven distribution across lifecycle stages reveals the community has invested considerably more in interpreting and checking documents than in supporting accurate and traceable generation.


\subsection{Open Science}~\label{appendix_openscience}
The venue list, analyzed papers, and the codebook are available at: \url{https://tinyurl.com/2c4exxd8}.

\subsection{Limitations}

We only searched through titles and abstracts during our initial steps of data collection. Thus, it is possible that some relevant works may have been missed. However, if our search terms were not present in this text, it is likely that privacy documents are not a primary focus of the paper. 

We also excluded several conferences and journals from our initial list of venues to make the data collection process more manageable. 
We aimed to mitigate the effects of this by performing an additional round of both forward and backward snowballing without the venue limitation. 
The lack of findings from an additional round of snowballing on a subset of these papers indicates that no significant papers were missed. 
Thus, we believe the vast majority of relevant works were captured.

Additionally, we did not calculate an agreement during the initial code-development stage, because inter-rater reliability may be inappropriate when codes are the object of iterative development~\cite{mcdonald2019reliability}. 
We instead supported consistency through independent pilot coding, consensus-based codebook construction, and collective discussion of edge cases
\section{The Format and Scope (RQ1)}
\label{rq1_def}

In this section, we examine the research target selection in prior studies, \textit{i.e.}, the format and scope of privacy documents.

\trend1 \textbf{Researchers mainly focus on privacy policies.}
Privacy policies represent the earliest and still most widespread form of privacy documentation. 
These long-form textual disclosures aim to comprehensively describe how personal data is collected, used, shared, and protected.
Although they serve as the legal backbone of modern privacy communication, their length, ambiguity, and inconsistent structure have drawn long-standing criticism and motivated subsequent innovations~\cite{amos2021privacy, shipp2020private, wagner2023privacy, adhikari2023evolution, singh2011evaluating, tao2025longitudinal, liu2016modeling, windl2022automating, pan2024hope}.
As expected, the privacy policy remains the dominant form of privacy documentation among existing research, appearing in 227 papers.
It illustrates the field’s longstanding emphasis on lengthy disclosures, as well as the relatively limited attention to more user-centric, interactive, or visual forms of communication.

Additionally, privacy policies are often complemented by more specialized disclosures, most notably cookie notices and cookie policies, with 14 relevant papers. 
While privacy policies describe data practices at global level, cookie-related disclosures focus on tracking technologies and consent mechanisms, particularly in web-based environments~\cite{matte2020cookie, bollinger2022automating, khandelwal2023automated, bouhoula2024automated, biselli2024supporting, jiwani2024crumbling}.

\trend{2} \textbf{The formats of privacy documents constantly evolve}, and greater attention should be placed on \opp{1} \textbf{emerging privacy document formats.}
Privacy labels provide concise summaries of a software's data practices, presenting key elements in a standardized and table-like format. 
These labels aim to reduce cognitive load, improve decision-making, and create industry-level consistency in privacy disclosures~\cite{pan2023toward}, which are mentioned in 28 papers, as the second most studied format.
Other forms appear less often with only a few mentions, such as contextual privacy policies~\cite{windl2022automating, pan2024hope, gong2025towards}, privacy icon~\cite{habib2021toggles}, runtime privacy notice~\cite{li2024we}, and data processing agreement~\cite{amaral2023ml}.
The design and evolution of privacy documents reflect ongoing attempts to enhance user comprehension, support automated processing, and align disclosures with increasingly complex data practices. 
These formats reflect an increasing effort to improve transparency and usability through visual and interactive design, and future research should place greater emphasis on these emerging formats.


\trend3 \textbf{Existing research center on english documents and dominate markets}, effectively lacking research effort on \opp{2} \textbf{other linguistic and cultural contexts.}
The majority of existing analyses focus on English-language privacy documents, with only few exceptions from other linguistic and cultural contexts, such as those in Germany~\cite{utz2019informed, arora2022tale, ganglmair2024regulatory, hosseini2025privacy, hosseini2024bilingual, freiberger2024fair, gambato2024effective}, China~\cite{liu2022evaluating, jiang2023personal}, India~\cite{de2022analysis}, and South Asia~\cite{javed2020study}. 
Broader inclusion of diverse languages and jurisdictions would enable a more comprehensive understanding of global privacy communication practices and how they interact with local cultural norms and legal frameworks.







\section{Privacy Document Generation (RQ2)}


This section examines how prior work supports the generation of privacy documents along two dimensions. 
First, we analyze generation approaches and the techniques applied. 
Briefly, privacy documents take various forms (\trend2), each with its own generation approach or pipeline. 
Privacy policies can be manually created by legal experts or via automated generators, such as questionnaires~\cite{zimmeck2021privacyflash, pan2023large}, templates~\cite{johnson2010optimizing, yu2016toward}, and code-based frameworks~\cite{li2024matcha, jain2023towards, yu2016toward, xiao2023enhancing, yu2015autoppg, jain2022creating}. 
Privacy labels can be generated by policy summarization and code analysis, like permission mapping~\cite{jain2023towards, jain2022pact, yu2015autoppg, yu2016toward, pan2023toward}. Second, we discuss the human factors in generation process, distinguishing intended users of these approaches. This allows us to identify both technical gaps and socio-technical gaps. 

\subsection{Approaches and Techniques}

\trend{4} \textbf{Generation is relatively less discussed}, and \trend{5} \textbf{existing research mainly focus on artifact-driven \textit{Code-based Generators} and \textit{Policy Summary Generation}.}

\textit{Code-Based Generators} refers to tools generating privacy documents leveraging program analysis and in-situ generators embedded in integrated development environments (IDEs) that assist developers~\cite{yu2016toward, yu2015autoppg, aydin2017visual, li2021toward, ke2023user, jain2022creating, jain2022pact, hjerppe2023extracting, xiao2023enhancing}.
In total, we identify 9 papers adopting this approach.
Those approaches infer privacy disclosures from software artifacts by extracting privacy-relevant evidence (e.g., personal data permissions, data-flow paths and other code behaviors) and translating it into aspects of privacy documents. 
IDE integration brings this pipeline directly into the coding workflow.
For instance, Matcha~\cite{li2024matcha} provides an IDE workflow that analyzes code to infer privacy disclosures to create privacy nutrition labels for developers. This approach addresses developers' limited legal and privacy expertise when creating privacy documents. Code-based approaches often pair program analysis 
~\cite{li2024matcha, jain2023towards, yu2016toward, hjerppe2023extracting, xiao2023enhancing, yu2015autoppg, jain2022creating, li2021toward, aydin2017visual} with template-based methods 
~\cite{pan2023large, leichtpripocog, emami2021informative, palka2025make, yu2016toward, johnson2010optimizing, tosch2024privacy, tao2025privacy, li2021toward} to structure disclosures or use machine learning techniques to classify extracted information into label fields or disclosure categories. Program-analysis-based generators aim to ground privacy disclosures in observable code behaviors using static and/or dynamic analysis. Prior works~\cite{yu2015autoppg, yu2016toward, li2021toward, jain2022creating} extract privacy-relevant code snippets and translate them into readable policy descriptions. Follow-up efforts such as
PriGen~\cite{jain2023towards} add richer code context and fine-grained localization.
Template-based methods use predefined templates~\cite{johnson2010optimizing}, questionnaires~\cite{pan2023large}, or boilerplate~\cite{tosch2024privacy} to avoid omissions.
Formal \& Standardization~\cite{palka2025make, pardo2019analysis, hjerppe2023extracting} techniques derive privacy documents from standardized representations. 
Learning-based methods further support generation by using neural models to classify privacy behaviors in Android code~\cite{jain2022pact} and translate code-level information into privacy captions~\cite{jain2023towards}.

\textit{Policy Summary Generation} digests out brief and clear summaries of lengthy privacy policies to improve user comprehension~\cite{zimmeck2014privee, fukushima2018challenges, harkous2018polisis, pardo2019analysis, emami2021informative, keymanesh2021privacy, brunotte2022my, hamid2023genaipabench, jain2023towards, pan2024hope, li2024matcha}, accounting half ofpapers within this RQ.
Summary generation is a common approach, likely because it directly targets the readability bottleneck posed by long, difficult-to-comprehend policies (\trend{12}). 
Summary generators~\cite{brunotte2022my, keymanesh2021privacy} are often applied to generate user-facing explanations, such as browser extensions for policy comprehension~\cite{brunotte2022my}. 

\opp3 \textbf{Authentic statement creation by contextual code evidence extraction.} 
Existing code-based approaches often rely on localized code evidence extraction, like isolated permission usages and limited code segments, which can result in generic privacy statements that do not fully reflect the actual privacy practices and do not always capture where the data ultimately goes, undermining downstream propagation or sharing contexts needed for accurate disclosures. 
A research direction could be to reconstruct the full lifecycle of personal information from source to sink. For example, PCapGen~\cite{jain2026automated} took the first step by leveraging taint analysis to recover complete data-flow paths and then using LLMs to produce more specific privacy captions. 

\textit{Organizational Generation Pipeline} treats privacy document generation as a cross-role workflow in which multiple organizational roles coordinate via shared artifacts. Five publications adopt this view~\cite{wishart2010collaborative, johnson2010optimizing, tosch2024privacy, leichtpripocog, tao2025privacy}.  PriBOM~\cite{tao2025privacy} proposes a privacy information inventory for collaborative notice generation in a development team, enabling roles 
to collaboratively populate and review the information, which can then be mapped to privacy notice clauses. Organizational software engineering practices are applied in these pipelines, considering lifecycle handoffs~\cite{tao2025privacy}. Auto update mechanisms are used to automatically update privacy policies when the underlying environment evolves~\cite{ke2023user}. Collaborative editing, where privacy documents are produced through multi-party co-authoring, supports negotiated and incremental updates~\cite{wishart2010collaborative}.

\textit{Online Automated Generators} refers to web-based automated tools for generating privacy documents through interactive forms or pre-defined templates. These generators typically guide developers through question-and-answer interfaces to produce general privacy policies that satisfy baseline legal requirements. 
Online Automated Generators are less discussed~\cite{pantrap, palka2025make}.
They commonly rely on questionnaire-driven pipelines and generate low-quality privacy policies for developers as online tools~\cite{pantrap}.

\opp4 \textbf{Standard formats for traceability and fine-grain generation.} Existing generators often rely on ad-hoc schemas, which makes it hard to maintain traceability links, e.g., which exact code segments justify a specific privacy statement, and support fine-grained updates. Inconsistencies and inaccuracies caused by the lack of traceability between code and privacy practice descriptions lead to inaccurate privacy notices~\cite{jain2022creating}. A future direction is to design and standardize a traceability record format where evidence pointers can be taint paths~\cite{jain2026automated} or localized privacy actions~\cite{jain2023towards, jain2022pact}. One effort towards this direction is PAcT~\cite{jain2022pact}, which establishes traceability from code behavior to privacy labels. PriGen~\cite{jain2023towards} further proposes fine-grained localization of privacy behaviors by locating code statements and using them to predict privacy labels.

\subsection{Human Factors in Generation Process}

Beyond the technical mechanism, aforementioned generation approaches also differ in whom they are designed to support.
The target audiences can be roughly categorized as \textit{non-technical roles},\textit{ citizen developers}, and \textit{developers in professional teams}.

\textit{Non-technical roles} include end-users, legal experts, regulators, and other stakeholders with distinct needs~\cite{palka2025make, zimmeck2014privee, pan2024hope, harkous2018polisis, hamid2023genaipabench, pardo2019analysis, emami2021informative, fukushima2018challenges, leichtpripocog, brunotte2022my, keymanesh2021privacy, wishart2010collaborative}. 
End-users require concise and plain-language explanations, whereas legal experts need complete and precise text with software feature context to draft disclosures. 
Regulators require structured and evidence-backed document to assess compliance.
Generation approaches targeting end-users often include summarization, customization, and visualization features.

\textit{Citizen developers} are developers outside formal teams who need to produce required privacy documents personally~\cite{li2024matcha, jain2023towards, jain2022pact, yu2016toward, hjerppe2023extracting, aydin2017visual, tosch2024privacy, pan2023large, yu2015autoppg, jain2022creating, li2021toward} . 
They often treat the privacy document generation as an one-time effort. 
Despite the target audience group, they assume software practitioners have a certain level of legal expertise, leaving terms such as \textit{``lawful bases''} and \textit{``Personally Identifiable Information (PII)''} for them to interpret, inevitably leading to misunderstandings. 

\opp5 \textbf{Under-evaluated usability from developers' perspective.} 
Evaluation of privacy document generation approaches often emphasizes technical correctness, such as whether a tool can correctly align a policy snippet about ``location data'' with relevant APIs, while neglecting usability in the engineering workflow, e.g., whether developer pain-points can really be reduced. 
Research on usability from the developer's perspective is limited. 
Usability questions, such as how much effort is needed to use and deploy the tool, whether developers can understand and act on its outputs, and how the workflow can fit into team processes, are often not evaluated or discussed.
One early effort in this direction is PriGen~\cite{jain2023towards}, which reportedly enabled less-experienced developers to reduce the time required to draft privacy statements by up to 74\%.
Onboarding and integrating new tools in practice remain a challenge~\cite{tao2025privacy}. Future work should discuss these usability dimensions explicitly and incorporate usability evaluations, measuring them alongside technical correctness to ensure that generated policies are not only correct on paper but also usable for developers in practice.

\trend{6} \textbf{Individual human-factors are commonly considered}, whereas \opp6 \textbf{the collaborative and dynamic nature of software development environments} is often overlooked.
\textit{Developers in professional teams} need to collaboratively develop and maintain software, meanwhile some of them handle the privacy disclosures across releases. 
We find that this group is least targeted, as only a relatively small group of publications~\cite{tao2025privacy, xiao2023enhancing, johnson2010optimizing, ke2023user} explicitly design organizational workflows to generate and maintain privacy documents in multi-role settings. 
For example, PriBOM~\cite{tao2025privacy} uses a shared inventory so stakeholders can collaboratively populate and maintain information for privacy document generation.
In practice, privacy disclosures should be developed collaboratively and co-evolve with the underlying codebase. 
PCapGen~\cite{jain2026automated}, for instance, combines heuristic taint analysis with LLM–based generation to produce traceable privacy captions based on the flow of personal data along taint paths.
We argue for a tighter integration of privacy documentation into DevOps practices, where multiple roles contribute and continuous maintenance and traceability are emphasized. Within such dynamic environments, updates to privacy documents should be systematically driven by code changes. 

\opp7 \textbf{The tension between development productivity and privacy responsibility.}
The emergence of generative AI has accelerated software development practices, enabling rapid prototyping, iterative deployment, and ``vibe coding'' approaches that prioritize speed and convenience. 
However, this increased productivity introduces a fundamental tension with privacy responsibility~\cite{pan2024don, zakharov2025ai}. 
Fast-paced development cycles may deprioritize careful consideration of privacy implications, resulting in privacy documents that are incomplete or outdated with actual software system behavior. 
When code and features are generated or modified rapidly without systematic privacy auditing or specialized legal expertise, the corresponding documentation may fail to accurately reflect actual software practices.
Future research could address this tension by integrating privacy-aware CI/CD pipelines, enabling co-generation of code and privacy disclosures, embedding compliance-by-design into development workflows, and empirically studying how AI-assisted practices influence developer behavior and privacy outcomes.


\section{Privacy Policy Content Analysis (RQ3)}
~\label{sec_rq3}

Privacy document content analysis constitutes a fundamental prerequisite for evaluating both compliance and usability, representing a critical stage in the privacy document lifecycle.
Broadly, existing methods for analyzing textual content can be categorized into rule-based and learning-based approaches.


\subsection{Rule-based Analysis}

\trend{7} \textbf{Rule-based analysis methodologies provide structured representations and constitute the foundation of privacy document analysis.}
Rule-based analysis methodologies can be broadly categorized into three main types:
\textit{Symbolic NLP}, \textit{Policy Formalization}, and \textit{Manual Coding-based Analysis} 
They commonly provide structured representations for organizing concepts expressed in privacy document texts, thereby facilitating systematic comparison across documents and enabling the identification of inconsistencies against software behaviors and noncompliance against regulations.

\textit{Symbolic NLP} methods are grounded in traditional NLP techniques, such as syntactic analysis and sentence parsing, typically involve parsing sentences into entities and subsequently extracting the underlying relationships among them. 
Our codebook identifies 77 publications (26.6\%) that use symbolic NLP techniques~\cite{zimmeck2014privee, andow2019policylint, yang2021purext, xiang2023policychecker}. 
A common theme in papers that make use of rule-based analyses is the development of ontologies that provide a consistent structure to organize concepts in policy texts, enabling more efficient analysis when comparing policies or identifying inconsistencies. 
PrivOnto (2018)~\cite{oltramari2018privonto} introduces a broad semantic framework for analysis, which can be used to generate a \emph{knowledge base} of annotated data practices contained within a privacy policy.
Stored data practices can be then retrieved by corresponding SPARQL queries. 
GUILeak (2018)~\cite{wang2018guileak} constructs a privacy policy lexicon from extracted information types and identified \emph{semantic relationships among phrases in the lexicon} to yield the ontology. 
PolicyLint (2019)~\cite{andow2019policylint} develops a set of ontologies to define \emph{subsumptive relationships} between terms in policies to allow detecting self-contradictory statements. 
PoliCheck (2020)~\cite{andow2020actions} can perform \emph{privacy data-flow analyses} by leveraging two ontologies for data types and entities. 
PoliGraph (2023)~\cite{cui2023poligraph}, is a type of \emph{knowledge graph} that captures statements in a privacy policy as relations among different parts of the text.

\textit{Policy Formalization} approaches transform privacy document text into a clear and unambiguous language, often a logic-based representation (\textit{a.k.a.} formal expressions), rather than parsing privacy document text based on their natural language properties. 
They are primarily used to standardize how the text is interpreted for developers or for automation purposes~\cite{breaux2008analyzing, jafari2014framework}.
The clear logic expressions can further facilitate anomaly detection and policy enforcement, especially in requirement engineering and software development~\cite{caramujo2019rsl, krasnashchok2020towards}.
Policy formalization approaches occurred less frequently, but was still explored in 24 publications (8.3\%).
A common theme of policy formalization is to concentrate on a particular aspect of the policy text, such as purpose~\cite{jafari2014framework}, information types~\cite{hosseini2021ambiguity}, consent~\cite{robol2023consent}, or software behaviors~\cite{yu2018enhancing}. 
This narrow scope highlights the inherent difficulty of achieving comprehensive and intrinsically rigorous formalization of privacy documents.
In particular, vague, conditional, and context-dependent statements, such as we may collect, ''adequate security measures'', or ``retain your data for lawful purposes'', are difficult to encode as precise formal expressions because their meaning often depends on legal, organizational, and contextual interpretation.

\opp{8} \textbf{Current feature extraction and standardization focus on data types, neglecting more subtle privacy practices.}
Most existing information extraction approaches concentrate on identifying explicit data practices, particularly the types of data being collected or processed (\textit{e.g.}, location, photos, and emails). 
However, this prioritization tends to underrepresent more nuanced and complex aspects of privacy practices, including third-party behaviors, user data rights, data security provisions, and mechanisms for risk and harm mitigation.
These dimensions are critical for a comprehensive understanding of privacy disclosures and often required by regulatory frameworks (\trend{10}), but remain insufficiently captured by current feature extraction schemas.
Closer collaboration with legal practitioners could support the development of more comprehensive and semantically grounded taxonomies, thereby fostering interdisciplinary approaches that better align technical analysis with legal and regulatory perspectives.
Furthermore, advances in LLMs present new opportunities to move beyond narrowly defined categories, enabling the large-scale extraction of richer and more contextual privacy practices.


\textit{Manual Coding-based Analysis} manually review policy documents strictly through coding or similar means, without substantial reliance on NLP or learning-based techniques.
In this context, the resulting codebook can be interpreted as a form of policy formalization, as it defines structured categories and criteria for analyzing policy content.
A significant portion (17.2\%) of the analyzed papers adopted this method~\cite{alomar2025effect, khandelwal2024unpacking, habib2019empirical, o2019reviewing, samarin2023lessons, manandhar2022smart, jesus2023feasibility}.
This approach is generally associated with smaller datasets, such as those targeting highly specific platforms or document sources, like Mastodon instances~\cite{tosch2024privacy} or South Asian websites~\cite{javed2020study}.

Given the subjective nature of manual annotation, the quality and credibility of results depend heavily on annotator expertise and agreement. Among the analyzed papers, 37 report annotator-related information, 
though only 12 explicitly provide inter-annotator agreement metrics such as Cohen’s Kappa. 
In terms of their expertise and background, approximately 15 studies involve PhD-level researchers, five involve graduate students, and four incorporate domain experts such as privacy lawyers.


\opp{9} \textbf{Human-in-The-Loop (HiTL) approaches.} 
Although the manual approaches are manageable for smaller datasets, more advanced NLP, especially LLM-based tools, could enable greater efficiency on privacy document analysis.
Mainstream LLMs are often pre-trained for generic NLP tasks, and they often show limited capabilities on domain-specific data~\cite{ling2025domain}, such as privacy documents~\cite{mori2025evaluating}. 
The complexity of privacy documents often necessitates expert interpretation; therefore, fully automated approaches are not always suitable for analysis tasks. 
HiTL (\textit{a.k.a.}, human-AI collaboration) methods can be used to mitigate misinterpretations and errors while still utilizing advanced automation approaches. 
Although the rise of LLMs has been incredibly beneficial for document analysis, hallucinations and misinformation remain common challenges that could be missed when outputs of automation tasks are not checked~\cite{huang2025survey}. 
Future work could therefore investigate how to incorporate fact-checking-oriented HiTL methods into privacy document research and evaluate their effectiveness in improving analytical reliability and reducing hallucinations.

\subsection{Learning-based Analysis}

\trend{8} \textbf{Learning-based analysis 
inherits the taxonomy from rule-based analysis.}
\trend{9} \textbf{These techniques have evolved alongside advances in NLP; however, limited explainability and hallucinations continue to challenge practitioners.}
Learning-based analysis approaches namely leverage machine learning, particularly NLP techniques, to analyze privacy documents at scale. 
Researchers adapt and refine general NLP methods to address the unique challenges posed by privacy documents, such as for understanding legal terminology~\cite{sleimi2018automated} and identifying disclosed data practices~\cite{xiao2024measuring}. 
Learning-based analyses are the most common form of analysis in our codebook, possibly driven by rapid advancements in this area in recent years. 
We identify 78 publications (26.9\%) that utilized this form of analysis~\cite{duc2024advancing, salvi2024privacychat, huang2024analyzing, ali2023honesty}.
The rapid advancements in ML have also driven the evolution of learning-based analysis methods, enabling more sophisticated, accurate, and scalable tools. 
Early attempts include the application of traditional machine learning methods, such as \emph{Support Vector Machine} or \emph{Liner Regression}, for document content classification~\cite{wilson2016creation}.
After, \emph{BERT-based models} are frequently used in learning-based analyses, as they can enable highly efficient classification of text~\cite{bui2021automated, wagner2023privacy, amaral2023ml}. 
More recently, \emph{LLMs} have been increasingly applied to assess regulatory compliance, including with frameworks such as the GDPR~\cite{duc2024advancing, mittal2024democratizing}. 
LLMs are also used to extract key information from policy texts, particularly elements related to regulatory requirements, to evaluate whether such requirements are satisfied~\cite{pan2023toward, aberkane2025factors, mittal2024democratizing}. 
For example, Pan et al.~\cite{pan2023toward} explored hybrid frameworks combining traditional machine learning with LLMs, as well as agentic LLM pipelines, to analyze privacy policies and generate corresponding privacy labels.
Similarly, Tang et al.~\cite{tang2023policygpt} proposed PolicyGPT, which leverages prompt engineering to perform privacy policy analysis simply using ChatGPT.   

\begin{figure}[t]
    \centering
    \includegraphics[width=\linewidth]{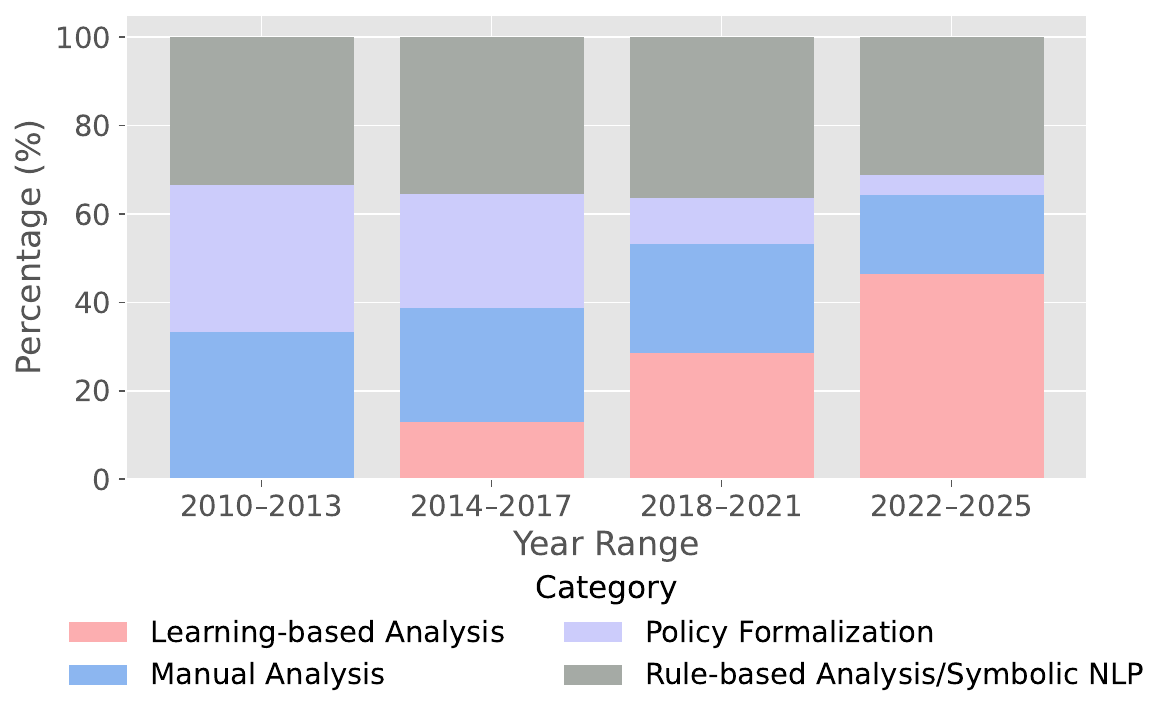}
\vspace{-15pt}
    \caption{Category distribution by year range (normalized).}
    \label{fig:RQ3_distribution}
\vspace{-15pt}
\end{figure}

\trend{10} \textbf{Manual analysis methods remain a major approach in privacy document analysis, despite the rise of learning-based methods}; and \opp{10} \textbf{large-scale expert-annotated manual analysis} becomes more essential. 
Learning-based analysis methods enable datasets to be labeled at a larger scale, whereas in the past, this work was conducted manually on smaller datasets. 
Fig.~\ref{fig:RQ3_distribution} shows the distribution of content analysis methods over the last 15 years.
Despite the increasing adoption of learning-based methods, manual analysis has remained one of the major approaches throughout this period.
Real-world privacy documents are often different from the generic documents that LLMs are typically trained on, as they might contain domain-specific legal terminology, complex clauses, and extensive cross-references (\trend{14}).
Moreover, the intrinsic nondeterminism of LLMs further undermines the trustworthiness of such black-box approaches.
Therefore, future work could aim to develop more tailored models using advanced approaches in prompt engineering, leveraging more sophisticated optimizations or rewards~\cite{cai2025unilaw}, and combining them with established rule-based analysis methods to enable stronger privacy-specific models.
A fundamental prerequisite for learning-based analysis is the availability of large-scale high-quality dataset, as model development typically requires substantial textual corpora. 
Consequently, the creation of large, expert-annotated privacy document datasets is increasingly critical for training, validating, and benchmarking learning-based approaches, particularly LLM-based ones.

\subsection{Dataset}


The choice of datasets across all analyzed papers varied widely. 
Some common datasets, like OPP-115~\cite{wilson2016creation}, were frequent, but many papers used self-collected datasets, requiring manual data collection.
This variety indicates that there is a growing need for an updated, large-scale, annotated corpus of privacy policies. 
Although manual collection has decreased as learning-based methods have become more popular, all future analysis would benefit from such corpora by eliminating unnecessary data collection stages. 
Future work could aim to develop and expand the existing datasets, such as PrivaSeer~\cite{srinath2020privacy} or the MAPS dataset~\cite{zimmeck2019maps} as a source for these documents.
In addition, emerging LLM-specific behaviors introduce new challenges for dataset design and analysis~\cite{tao2025longitudinal}. 
For example, the role of prompts and chat history as central inputs to model behavior necessitates necessitating datasets that explicitly capture interaction context, rather than simply treating them as monolithic categories (e.g., ``chat history'').

\opp{11}
\textbf{Longitudinal datasets for privacy document evolution.}
Current research on how to effectively address change management is limited. 
As software is updated, privacy documents can quickly become outdated and inaccurate. 
More work here could better ensure that documents remain up to date when software behavior is modified. 
Likewise, as regulations are updated, developers also need help identifying non-compliant text and the resulting necessary changes. 
Future work could develop approaches that automatically identify changes in software or regulatory documents and re-evaluate policies accordingly. 
These approaches could also implement traceability features to present a clear history of document changes (\opp{3} and \opp{4}), such that changes are always intentional and transparent. 

\opp{12} \textbf{Multilingual and multi-source datasets.}
More exploration is also needed in the area of multilingual privacy document analysis and dataset construction. 
With the majority of documents written in English, analysis of non-English policies is very infrequent. Although some papers~\cite{ur2012privacy, arora2022tale, hosseini2025privacy} aim to analyze non-English documents, such as their differences with their English counterpart, there is a limited availability of tools and frameworks outside of English-speaking contexts. 
Some existing annotation frameworks may be applicable to non-English policies, but it is unclear whether the same degree of analysis can be performed without additional work.
In addition, existing datasets primarily emphasize \textit{vertical}, single-document analysis, often neglecting \textit{horizontal} analyses across multiple sources.
For instance, there is limited work examining the alignment and consistency of privacy policies across different platforms and ecosystems, such as Google Play, iOS, Samsung Galaxy Store, and Huawei AppGallery.
Strengthening international collaboration will also be critical to constructing diverse, representative, and high-quality datasets for global privacy research.
\section{Consistency and Compliance Checking (RQ4)}
\label{rq4}

Compliance reflects how regulatory bodies interpret and evaluate privacy documents in relation to legal frameworks and observed system practices, representing a central function in their lifecycle.
However, from a legal and formal perspective, compliance is inherently difficult to define and assess in a uniform manner. 
Prior work frequently uses terms such as ``(non)compliance'', ``(in)consistency'', ``discrepancy'', ``(mis)alignment'', ``(mis)matching'', and ``violation'' interchangeably, despite their potentially distinct meanings depending on the context and scope. 
For clarity and terminological consistency, we adopt the term ``(non)compliance'' when referring specifically to relationships with regulatory requirements, and ``(in)consistency'' for other contexts.

In this section, we summarize existing studies across multiple five dimensions, including inconsistencies between software behavior and privacy policies (\textit{software–policy}), noncompliance between policies and regulatory requirements (\textit{policy–regulation}), and noncompliance between software behavior and regulatory expectations (\textit{software–regulation}). 
We also consider \textit{inconsistencies across sources} (e.g., privacy policies and privacy labels), as well as \textit{self-contradictions} within individual privacy documents.

\subsection{Software-Policy Inconsistency}

\trend{11} \textbf{Software-Policy inconsistency studies mainly focus on mobile applications, driven by the availability of mature program analysis frameworks.}
The consistency between software and privacy policies is the foundation of privacy compliance of the whole software ecosystem~\cite{andow2020actions, xiao2023lalaine}, and we identified 62 works targeting on this topic.

Mobile applications, particularly Android apps, are a primary focus of this form of consistency analysis due to their widespread use, Android's open-source platform, and the well-designed Android permission mechanism, which provides a structured framework for monitoring data access and usage~\cite{zimmeck2019maps, zimmeck2021privacyflash}.  
Hashmi et al.~\cite{hashmi2021longitudinal} investigated these inconsistencies in Android applications over time, finding that software behaviors inconsistent with their respective privacy policies have increased over time. 
Specific types of apps are also commonly evaluated, such as AI chatbot applications. 
Ragab et al.~\cite{ragab2024trust} analyzed romantic AI chatbot apps for inconsistencies between privacy policies and chatbot responses to privacy-related questions. 
In contrast to open-source Android apps, only three papers~\cite{bui2021consistency, xiao2023lalaine} have investigated the closed-source iOS apps.

Websites are also commonly analyzed for these inconsistency instances, typically focusing on consent,  data control implementations, and third-party data collection~\cite{libert2018automated, bouhoula2024automated}.
Most website-focused studies rely on dynamic analysis techniques, such as inspecting network traffic and web tracking behaviors, to determine whether implemented data practices align with privacy policy disclosures.
Other platforms, like virtual reality (VR), have also been evaluated.
Trimananda et al.~\cite{trimananda2022ovrseen} introduced OVRSEEN, a methodology for analyzing privacy risks in the Oculus VR ecosystem by examining network traffic and privacy policies. 
They find that about 70\% of data flows were not properly disclosed. 
Across these diverse platforms, inconsistencies are found to be commonly present.

\opp{13} \textbf{Program analysis framework for emerging software formats and AI-centric ecosystems.}
A key challenge in such consistency checking is establishing a reliable connection between specific privacy disclosures and respective software behaviors. 
Studies often propose frameworks that leverage both static and dynamic analysis to reveal privacy practices and identify inconsistencies between app behaviors and policies~\cite{wang2025miniscope, yan2024investigating, tan2023ptpdroid, trimananda2022ovrseen, hashmi2021longitudinal, hatamian2021privacy}. 
For example, Wang et al.~\cite{wang2018guileak} mapped Android GUI input views to policy terms using an ontology and applied static information flow analysis to detect privacy leaks. 
Similiarly, Andow et al.~\cite{andow2020actions} proposed POLICHECK, an entity-sensitive flow-to-policy consistency analysis tool that identifies privacy-sensitive data disclosure violations in mobile applications, enhancing transparency and accountability across app data practices as a whole.
In contrast, analyzing software in other forms, such as iOS apps, desktop software, and AI-based applications, encounters greater challenges due to their more closed-source architectures or lack of standardized program analysis frameworks.

\subsection{Policy-vs-Regulation Noncompliance} 
\trend{12} \textbf{Existing studies mainly focus on mainstream regulations, namely GDPR, with scrutiny largely centered on the under-disclosure of collected data}; \opp{14} \textbf{emerging regulations (e.g., the EU AI Act) and broader jurisdictional coverage} remain underexplored (\trend{3}).
Compliance checking between privacy policies and regulations involves determining if statements in privacy policies adequately address the requirements set by applicable laws. 
A common form of noncompliance is under-disclosure, where required data collection behaviors or data rights are absent from privacy policies.
These omissions undermine regulatory compliance and potentially lead to serious legal consequences.
The majority of research focuses on exploring GDPR compliance. 
For example, Xiang et al.~\cite{xiang2023policychecker} designed PolicyChecker, a framework to assess GDPR compliance in mobile app privacy policies by evaluating their completeness against all mandatory and conditional GDPR requirements. 
Using a rule-based and semantic role analysis approach, PolicyChecker systematically identifies GDPR violations, highlighting gaps in data collection disclosures and user rights information.

Some work has also explored how policies under GDPR compare to those in other regions, like the CCPA. 
Arora et al.~\cite{arora2022tale} investigated how privacy policies differ under two major regulatory regimes, the EU’s GDPR~\cite{GDPR} and California’s CCPA~\cite{CCPA}. 
The authors created a bilingual corpus of privacy policies and trained machine learning models to automatically extract disclosures from privacy policies, revealing that GDPR leads to stricter privacy disclosures. 
A few works focus on compliance checking strictly for regulations in non-EU countries. 
Mori et al.~\cite{mori2023analysis} examined the privacy policies of Japanese companies to evaluate their compliance with Japan's Act~\cite{japan_appi_2003} on the protection of personal information. 
They find that the vast majority of companies fail to comply with data retention requirements. 
Tocchini et al.~\cite{tocchini2024classifying} explored Portuguese-language sentences for compliance with Brazilian privacy regulation~\cite{brazil_lgpd_2018} through the use of classification models. 
These works indicate that privacy violations are also present in non-English contexts, which is still underexplored.

\opp{15} \textbf{Regulatory ambiguity and context-dependent interpretation pose challenges for technical researchers.}
Existing research on noncompliance often treats regulatory documents as objective and unambiguous standards against which privacy policies can be directly evaluated~\cite{gupta2022creation}. 
However, in practice, regulatory texts are inherently ambiguous due to the nature of languages, posing significant challenges for researchers, particularly those from technical backgrounds.
Ambiguities in legal provisions can lead to divergent interpretations, and regulatory requirements are often highly context-dependent. 
As a result, current studies tend to focus on clear-cut and less controversial cases of noncompliance, such as under-disclosure of data practices.
Interdisciplinary collaboration between legal scholars and computer scientists will be essential to ensure more accurate and context-aware interpretations of regulatory requirements.
In addition, LLMs may provide support in interpreting complex legal text; however, their application must be carefully validated given commonly-criticized hallucinations. 
Furthermore, in jurisdictions governed by common law, prior case rulings offer valuable precedents for interpreting ambiguous provisions and could be leveraged to enable more nuanced and context-aware compliance analysis.



\opp{16} \textbf{Compliance with industry standards and platform-specific requirements.}
Existing research on privacy compliance focuses on formal regulatory legal frameworks, while largely overlooking industry standards (e.g., ISO standards~\cite{ISO_InfoSec} and NIST~\cite{NIST_PF}) and platform-specific requirements (e.g., Google Play Console specifications~\cite{googledeveloper}).
Notably, earlier work had actively explored such practical frameworks~\cite{guttman1995introduction, spiekermann2008engineering}, suggesting a shift away from operational compliance perspectives in recent research.
These forms of compliance are often more directly actionable and operationalizable, as they provide concrete and enforceable guidelines for developers and organizations.
Future work should therefore expand the scope of analysis to include these practical compliance frameworks, investigating how they can be systematically integrated into existing methodologies. 

\subsection{Software-vs-Regulation Noncompliance} 

This category refers to software behaviors that are non-compliant with privacy regulations.
We identify a total of 15 works that aim to correlate specific software behaviors with violations of regulatory requirements, either in isolation or alongside with software-policy consistency analysis. 
In this context, privacy policy analysis often serves as a bridge between observed software behaviors and legal requirements.
Some studies focus directly on consent mechanisms.
For example, Nguyen et al.~\cite{nguyen2021share} explored consent mechanisms required by GDPR in Android applications, finding that 24,838 sent data to data controllers without explicit user consent. 
Iwaya et al.~\cite{iwaya2023privacy} examined mental health apps for common privacy principles, finding significant threats to information disclosure as well as insufficient notice mechanisms.


\opp{17} \textbf{Real-time compliance checking for adaptive and rapidly evolving software.}
Modern software systems are increasingly adaptive and rapidly evolve in the AI era.
Emerging paradigms, such as LLM-based agents and modular ``skills'', enable rapid prototyping, dynamic behavior adaptation, and continuous deployment, potentially being applied in high-stakes application domains.
As software is increasingly developed through ``vibe coding'' approaches, often by citizen developers, software behaviors may unintentionally fall out of compliance. 
While existing research has explored compliance checking in a post-hoc manner, there is a growing need for approaches that ensure continuous compliance through real-time verification mechanisms~\cite{chowdhury2014temporal}.
Future work should therefore explore the development of automated tools that monitor compliance throughout the software lifecycle.

\subsection{Inconsistency between Sources} 

\trend{13} \textbf{Inconsistencies across sources are prevalent, such as mismatches between privacy policies and privacy labels.}
Different representations of privacy information for the same software should be consistent. 
However, in practice, inconsistencies are common due to mismatches in granularity, differing taxonomies, and inaccurate or incomplete disclosures~\cite{jain2023atlas, ali2023honesty, cranor2022mobile}; and relatively few studies have systematically examined such inconsistencies. 
Additionally, privacy documents are often presented in global markets where multilingual versions are required, and significant differences may affect fairness, transparency, and regulatory compliance. 

Several works examine inconsistencies in the information conveyed between privacy policies and their associated privacy labels~\cite{jain2023atlas, ali2023honesty, lin2023data}. 
These studies typically apply the privacy policy analysis techniques discussed in Section~\ref{sec_rq3} to extract policy disclosures and map them onto privacy label taxonomies, especially the Google Play Data Safety~\cite{googledatasafety} and Apple App Privacy~\cite{appledeveloper}.
For example, Jain et al.~\cite{jain2023atlas} presented Automated Privacy Label Analysis System, a large-scale system that detects discrepancies between privacy policies and privacy labels in iOS applications. 
Similarly, Ali et al.~\cite{ali2023honesty} investigated the accuracy of Apple's privacy labels by accessing the discrepancies between the privacy labels of 474,669 iOS apps and the corresponding privacy policy features extracted by PrivBERT~\cite{srinath2020privacy}. 
The findings reveal that 97\% of iOS apps claiming ``Data Not Collected'' actually indicate data collection in their privacy policies.

\opp{18} \textbf{Exploration of multilateral inconsistency.} 
Existing studies on inconsistency focus on bilateral comparisons (i.e., between two sources), such as policy–label mismatches. 
However, in practice, privacy disclosures often span multiple interconnected sources, making multilateral inconsistency a more pressing and complex challenge.
For example, Pan et al.~\cite{pan2025look} conducted a multi-source consistency analysis of task-executable voice assistant applications, examining five distinct privacy disclosure sources: (1) privacy labels, (2) permissions identified by analysis tools, (3) permissions listed in mobile system settings, (4) privacy policies, and (5) declarations in manifest files. Their findings highlight the complexity of aligning disclosures across these heterogeneous sources.
Future work should therefore move beyond bilateral analysis, constructing datasets that capture these richer, interconnected representations (\opp{12}) and developing methodologies for systematic multilateral consistency checking.

\subsection{Self-contradictions} 
Even if documents are consistent with other sources, they can still contain internal contradictions, leading to confusion and potential noncompliance. 
Few works focused on such self-contradictions that exists within the same privacy document. 
We identify ten works that highlight these inconsistencies within a single document. 
The most detailed example of this analysis is in PolicyLint~\cite{andow2019policylint}, which develops a set of ontologies to detect contradictions in apps on the Google Play store. They find that of the 11,430 policies analyzed, 14.2\% contain some contradiction. 

\section{Usability and Readability (RQ5)}
\label{rq5_usability}

End-user consumption represents another primary objective of privacy documents, enabling them to understand how their personal data are collected, used, and shared, thereby supporting informed decision-making.
The usability and readability of these documents are therefore central to their effectiveness as mechanisms for transparency and accountability. 

\subsection{Usability Issues}

\trend{14} \textbf{Privacy documents are widely suffer usability and readability issues, they are commonly evaluated using metrics such as length, Flesch Reading Ease Test scores, and measures of vagueness.}
Across the literature, several recurring usability issues have been identified. 
The most frequently reported problem is vagueness (23 out of 126 usability papers), in which privacy documents rely on ambiguous or non-lay language that prevents users from forming a clear understanding of actual data practices.
To quantify vagueness, prior work frequently relies on the scores of the Flesch Reading Ease Test~\cite{kincaid1975derivation}.
Excessive length (24/126) is also a noticeable problem, which makes policies burdensome to read and discourages engagement altogether. 
The average lengths (by word count) of privacy policies reported by literature are 1,522~\cite{amos2021privacy}, 1,871~\cite{srinath2021privacy}, 2,500~\cite{adhikari2023evolution}, and 3,057~\cite{Javed2024}, with length substantially increasing over time~\cite{Javed2024, amos2021privacy}. 
Although most of the reported issues exist in privacy policies, emerging privacy documents, such as privacy labels, also suffer from vagueness~\cite{gluck2016short, garg2022impact, zhang2022usable, li2022understanding1, khandelwal2024unpacking, zhang2024exploring}.
Other commonly reported issues include privacy documents being difficult to find, for example, buried behind multiple navigation layers~\cite{hosseini2021unifying, audich2021improving, im2023less, mhaidli2023researchers}, and the complete absence of privacy documents in contexts where users reasonably expect them~\cite{shipp2020private, linden2018privacy, pan2024large}.
Together, these issues indicate that usability failures are not limited to linguistic complexity alone, but also stem from structural, contextual, and discoverability problems. 


\subsection{Usable Solutions}

\trend{15} \textbf{Privacy disclosures are increasingly designed to be just-in-time, contextualized, and more readable.}
Many work proposes solutions to improve the usability and readability of privacy documents, which can be roughly clustered into four recurring design paradigms.

\textit{Label-ization} aims to compress complex privacy disclosures into standardized, visual formats that enable quick comparison and rapid comprehension. 
Kelley et al.~\cite{kelley2009nutrition, kelley2010standardizing} introduced privacy nutrition labels as a structured, tabular representation of key privacy practices, inspired by food nutrition labels, demonstrating improved comprehension speed and reading experience compared to full-text policies. Subsequent work extended this concept to mobile applications~\cite{kelley2013privacy} and IoT devices~\cite{emami2020ask}, supporting installation and purchasing decisions in constrained interfaces.
Industry adoption of label-like disclosures, such as Apple App Privacy~\cite{appleprivacy} and Google Play Data Safety~\cite{googledatasafety}, further illustrates the scalability and practical impact of this paradigm.

\textit{Machine-Readable Formats} represents privacy disclosures in machine-readable and formally structured formats, enabling automated retrieval, analysis, and personalization. 
One of the earliest and most influential efforts in this direction is the Platform for Privacy Preferences (P3P).
The P3P enabled websites to express their privacy practices in a standardized, machine-interpretable format, alongside a human-readable version~\cite{cranor2002web}. 
This dual representation allowed user agents to automatically retrieve and interpret privacy statements, lowering the burden on users to manually read full policies. Despite its conceptual promise, P3P ultimately failed to thrive due to insufficient adoption by major browser vendors, leading to the suspension of further development~\cite{P3P}.
Subsequent research on privacy policy languages and formal representations~\cite{kumaraguru2007survey} continues this direction, emphasizing interoperability, traceability, and automated compliance checking. 
While these approaches improve scalability and precision, their benefits to end-users are often indirect.

\opp{19} \textbf{Machine-readable formats of personal privacy preferences/settings gain renewed relevance in the era of LLM agents and skills.}
The growing prevalence of AI-based agents, capable of autonomously browsing, querying, and acting on behalf of users, creates new opportunities for representing personal privacy preferences in structured, machine-readable formats. 
In such modern software ecosystems, LLM-based agents and modular ``skills'' could directly consume and operationalize these representations to support automated consent-related decision-making.
This shift highlights the renewed relevance of earlier efforts, such as the P3P, facilitating more scalable, personalized, and automated privacy management within agent-based environments.

\textit{Contextualization} approaches aim to bridge the gap between textual policy text and the concrete situations in which data practices actually matter to users. 
Contextualization shifts privacy communication from static documents to more real-time disclosures, presenting relevant privacy information when it becomes actionable. 
Windl et al.~\cite{windl2022automating} proposed contextual privacy policies that embed relevant disclosures directly into website interactions, aligning privacy information with concrete user actions, and Pan et al.~\cite{pan2024hope, gong2025towards, tao2026contextual} extended it to mobile platforms. 
Building on this idea, contextual approaches have been extended to AI systems. 
For example, CLEAR provides just-in-time privacy explanations and risk highlights during user interactions with LLM-powered applications~\cite{chen2024clear}.
By grounding privacy statements in specific contexts, these approaches reduce abstraction and support situated decision-making. However, they introduce design challenges related to timing, interruption, and cognitive overload.

\textit{Personal Privacy Assistants} (PPAs) conceptualize usability as ongoing, personalized interaction rather than one-time disclosure. 
Instead of expecting users to read, compare, or interpret privacy documents themselves, PPAs act as intermediaries that retrieve, interpret, and contextualize privacy information on demand.
Prior work on document-centric privacy question answering (Q\&A) has laid a foundation for meaningful PPAs~\cite{colnago2020informing, keymanesh2021privacy, hamid2023genaipabench}. 
Recent advances in LLMs further expand the potential of PPAs by enabling more flexible reasoning, synthesis, and personalization~\cite{pan2024hope, chen2024clear}.
However, the inherent privacy and security risks of AI systems, along with their opacity and potential for hallucination, may undermine the reliability and trustworthiness of AI-empowered PPA without careful handling.
Future research should explicitly focus on trustworthy, verifiable PPAs, such as grounding PPA responses in traceable evidence from privacy documents, software behaviors, or regulatory sources.

\opp{20} \textbf{Usability solutions remain system-centric rather than user-centric.}
Many existing usable solutions remain implicitly grounded in system- or development-centric abstractions. 
For example, in the Android ecosystem, privacy documents are often derived from the permission-related API usages and manifest declarations. 
Also, the compliance scrutinization of relevant documents is heavily rooted in the Android permission systems. 
While such infrastructure-driven approaches improve technical consistency, they do not necessarily align with how users perceive or reason about privacy risks. 
Future usability improvements could be grounded in users’ mental models, situational concerns, and perceived consequences, rather than mirroring system-level primitives.
For example, disclosures could highlight concrete negative outcomes, such as tracking their location in the background, more targeted ads, or risks if data are breached~\cite{ebert2026procedures}.

\opp{21} \textbf{Actionable privacy control mechanisms.}
Current privacy document research largely emphasizes \emph{notice} over \emph{control}, stating users’ rights in abstract legal terms rather than providing actionable instructions. 
For instance, explicit instructions within a privacy center on how to delete collected data or adjust privacy settings (e.g., ``Settings $\rightarrow$ Account $\rightarrow$ Collected Data Management $\rightarrow$ Data Entry Deletion'') are often more understandable and practically useful than statements such as ``\textit{... you are entitled to the right to be forgotten.}'' 
Integrating privacy controls and step-by-step instructions directly into privacy documents is an important direction for improving usability. 
Although early industry trends toward centralized privacy dashboards and control hubs suggest growing momentum in this direction, systematic academic investigation of their design, effectiveness, and user impact remains limited.


\section{Identified Research Directions}
\label{sec:research_framework}

\dir{1} \textbf{AI-centric platforms and their emerging privacy challenges.}
While existing studies have extensively examined privacy documents within traditional software such as websites and mobile apps, the rapid emergence of AI-centric platforms, such as LLM-based systems, agents, and modular ``skills'', is fundamentally reshaping how software is developed and deployed, as well as how personal data is collected and processed.
This shift responds to the limited attention paid to emerging software formats and AI-centric ecosystems (\opp{13}), and further intensifies the need for real-time compliance checking in rapidly evolving software (\opp{17}). First, AI systems are inherently interaction-centric, where prompts and conversational histories serve as both inputs and evolving context.
This challenges conventional notions of data collection and storage, as privacy-relevant information is no longer confined to predefined data fields (e.g., permissions and APIs in Android apps), but dynamically generated through user interactions.
Second, these systems exhibit increasing levels of autonomy. Agent-based architectures can act on behalf of users, including performing complex and potentially sensitive operations across services.
This raises new questions regarding consent and transparency, and how such delegated actions should be represented within privacy documentation. 
These developments also renew the relevance of machine-readable privacy preferences and settings (\opp{19})

\dir{2} \textbf{Update-to-date and diverse data foundations.}
Foundational datasets, such as those collected and maintained in Usable Privacy Policy Project~\cite{Usable_PP_Project}, have played a critical role in advancing privacy document research. However, there remains a need for updated and more comprehensive data infrastructures that better reflect the evolving landscape of privacy practices.
Across all RQs, a recurring bottleneck is the lack of shared, high-quality datasets that capture the diversity of privacy documents in real-world settings. 
Most existing datasets are text-centric, English-only, and primarily focused on long-form privacy policies, limiting their applicability to modern, heterogeneous privacy ecosystems and emerging privacy document formats (\opp{1}). 
Addressing this limitation requires a shift toward richer and more representative data foundations.
In addition to longitudinal datasets (\opp{11}), multilingual and multi-source coverage (\opp{12}), future efforts should further expand to multimodal datasets that beyond textual privacy policies to include multiple modalities, such as privacy labels, privacy icons, user interface contexts, and graphical screenshots.
Such datasets should also support large-scale expert-annotated analysis (\opp{10}), while enabling systematic study of multilateral inconsistency across interconnected sources (\opp{18}).
Capturing relationships across modalities is essential for enabling research on cross-format consistency and user-centric privacy disclosure design.

%

\dir{3} \textbf{LLM-based unified policy-code analysis.} LLMs have considerable potential for widespread adoption in privacy policy analysis, because of their strong capabilities in natural language understanding and their ability to capture nuanced semantic meanings beyond rigid predefined taxonomies~\cite{duc2024advancing, salvi2024privacychat, huang2024analyzing}.
This capability is particularly relevant to extracting more subtle privacy practices that are often neglected by current feature extraction and standardization efforts (\opp{8}).
However, the hallucinations and probabilistic nature have raised concerns regarding reliability and trustworthiness in high-stakes applications such as compliance checking, reinforcing the need for Human-in-the-Loop approaches (\opp{9}) and careful handling of regulatory ambiguity (\opp{15}).
Recent advancements, including retrieval-augmented generation and harness engineering, provide promising directions for improving LLM-based analysis. 
By grounding model outputs in verifiable evidence, these techniques can mitigate hallucinations and enhance the reliability of outputs.

Beyond standalone document analysis, a key emerging direction is the integration of LLMs with software artifacts. 
Joint analysis of privacy policies and corresponding source code can enable more accurate interpretation of policy statements by grounding them in observable system behaviors. 
This direction connects to authentic statement creation through contextual code evidence extraction (\opp{3}) and standard formats for traceability and fine-grained generation (\opp{4}).
Furthermore, combining LLM-based reasoning with dynamic analysis can provide additional empirical evidence to validate or challenge policy claims.  
For example, automated agents execute test cases to probe software functionality.
Such hybrid approaches can improve the reliability of privacy policy analysis, with more contextualized and faithful interpretation.

\dir{4} \textbf{Dual usability: end-users and developers.}
An evident imbalance persists in usability research, which focuses on end-users while largely neglecting the usability challenges faced by developers (\opp{5}). 
Nevertheless, developers encounter similar significant difficulties in interpreting and implementing privacy-related content.
Importantly, these two dimensions are interdependent. 
Poor developer-facing tools often lead to inaccurate or misleading user-facing disclosures. 
Evaluations should therefore consider bidirectional usability impacts rather than treating these audiences independently, as existing usability solutions often remain system-centric rather than user-centric (\opp{20}). 
Early efforts have focused on improving the accuracy of privacy disclosures during development, such as Honeysuckle~\cite{li2021honeysuckle} and Matcha~\cite{li2024matcha}.
Beyond technical correctness, usability metrics should capture effort, learnability, error prevention, and integration in collaborative and dynamic real-world software engineering practices, such as CI/CD pipelines, code review, and release management (\opp{6}).

Furthermore, evaluation frameworks should capture usability failures in developer workflows, such as situations where developers lack guidance or confidence in writing privacy documents and resort to copy-pasting boilerplate templates; when users recognize such generic or irrelevant disclosures, they may disengage entirely from privacy documents. 
Conversely, evolving user needs might introduce additional challenges for developers in accurately disclosing privacy practices, especially when they depend on third-party libraries or AI-based services whose data practices are externally controlled. 
Addressing these challenges requires across-role studies, and evaluating how effectively requirements, evidence, and responsibilities are communicated across organizational boundaries.
In addition, bidirectional evaluation should consider whether privacy documents can connect notice with actionable privacy control mechanisms (\opp{21}).

\section{Conclusion}


Over the past decades, research on privacy policies and documents has expanded significantly.
This SoK systematically organizes 290 papers published between 2010 and 2025 to provide a unified, lifecycle-oriented view of privacy documents from an engineering perspective.
Specifically, we examine how privacy documents are defined and scoped, created, analyzed and extracted, assessed for inconsistency and noncompliance, and evaluated and improved for usability.
Across these lifecycle stages, we summarize 15 research trends, identify 21 open opportunities, and chart four broader research directions.
By synthesizing prior work from a lifecycle perspective, we hope this SoK serves as a shared foundation for future research on privacy documents and supports the development of privacy communication mechanisms.


\section*{Acknowledgement}
The authors thank the participants of Dagstuhl Seminar 25021 ``Grand Challenges for Research on Privacy Documents'' for their insightful discussions. 
This work was supported in part by NSF CAREER Award \#2238047, NSF SHF Award \#2442683, the Seed Funds Program of Columbia University's Data Science Institute, and the Software Engineering for AI Focus Group under a Hans Fischer Senior Fellowship at the Technical University of Munich.

\IEEEpeerreviewmaketitle

\bibliographystyle{IEEEtran}
\bibliography{ref}


\end{document}